# Oscillator Chain: A Simple Model for Universal Description of Excitation of Waveguiding Modes in Thin Films

Kestutis Staliunas[1,2,3,4]

[1]ICREA, Passeig Lluís Companys 23, 08010, Barcelona, Spain
[2]UPC, Dep. de Fisica, Rambla Sant Nebridi 22, 08222, Terrassa (Barcelona) Spain
[3]Vilnius University, Faculty of Physics, Laser Research Center, Sauletekio Ave. 10, Vilnius, Lithuania
[4]Center for Physical Sciences and Technology, Savanoriu Ave. 231, LT-02300, Vilnius, Lithuania

**Abstract**

There is no simple and universal analytical description of various micro-optical systems related with Fano resonances. This especially concern modulated thin films, which, when coupled to external fields, show Fano resonances. Usually, such micro-optic circuits are simulated numerically, frequently by the use of commercial software. We fill this gap of the lack of universal analytical description by introducing and exploring a simple mechanical equivalent, the oscillator chain, which mimic such schemes involving Fano resonances. The model does not necessary provide the rigorous description of complicated micro-optical schemes, however does capture the main properties of such Fano-related micro-optical systems.

The model captures different modifications of the thin film arrangement as well: thin film with amplification, non-Hermitical thin films, and others. It also covers the case of multiple Fano resonances in a thin film. The model is compared with the rigorously calculated (Rigorous Coupled Wave analysis) wave propagation in the thin films.

## I. Introduction

The problem of a coupling of a continuous wave with an oscillator arises in many fields of physics. The first studies were performed by Struth (Lord Rayleigh), Mie, and others [1-3] to calculate the scattering of electromagnetic wave by a particle. A general formula to calculate the scattering cross-section of a particle resonantly interacting with the plane wave has been derived by Fano, [4,5]. The related problem is being encountered quite frequently now in modern micro/nano photonics, in particular in the studies of wave propagation in micro-circuits and interaction with micro-nano rings [6-15], or in spatial filtering by photonic crystals [16-18]. This problem is also encountered in the series of recent studies of the light reflection/transmission through the surface-modulated thin films in different modifications [19-24], Fig.1.a. These studies largely rely on numerical methods. The analytical studies are complicated, since even the simplest cases depend on many parameters and specifics of architectures, which neither allow to derive simple analytical models for particular configuration, nor to derive universal models. A derivation of a simple universal relation, even an approximate one, which would capture the main properties of such an interaction, would be very useful to develop an intuition of the waves propagating/interacting in such system, and would allow to predict the expected phenomena.

An analytical attempt to solve the problem is the coupled mode theory (CMT) [25-27], however, due to different possible configurations of the thin film modulation (for instance only one interface can be modulated, or both, with equal or different amplitudes), this also does not lead to a simple and universal expressions.

The original theory of Fano concerns the weak scattering of a plane wave by a particle, where the first-Born approximation [28] is usually applied. This means that the amplitude of incident plane wave is fixed, and the scattered field by the particle is considered as a weak perturbation, not altering the incident field in this fiorst approximation. The classical [1-5] and modern [6-15] studies calculates the scattering cross-section by a particle at- or near to resonance. The case in Fig.1.a is, however, principally different, in that the scattered field is comparable in magnitude with the incident wave, and can substantially affect it. The interference between the incident and scattered field can result in drastic changes in the domain of transmitted/reflected radiation. The usual scattering theory, based on first Born approximation, does not apply here.

The aim of this article is to derive a simple and universal model capturing the main features of the systems in Fig.1.a. The purpose is to correctly capture the extreme cases, when, for instance, reflection or transmission becomes zero, or increases to infinity in the case of the active (amplifying) thin film. Such cases usually are not accessible by perturbation theories, for instance by first Born approximations.

The situation is universally described by a simple model of the chain of oscillators coupled to the neighboring ones, and to additional oscillator say at a position, $n = 0$, see Fig.1.b. Along the chain at $n < 0$ the incident and the reflected waves are propagating. At $n > 0$ the chain contains only the transmitted wave. The oscillator at $n = 0$ mimics the Fabry-Perrot resonator for the field resonating nearly perpendicularly to the surface of the film in Fig.1.a. The oscillator B, coupled to the oscillator at $n = 0$ position of chain, corresponds to the planar waveguiding (or, in limiting case, the leaky surface-) mode of the film. Although very schematically, such chain of oscillators contains all the ingredients of the periodically modulated thin film coupled to the incident plane waves, Fig.1.a.

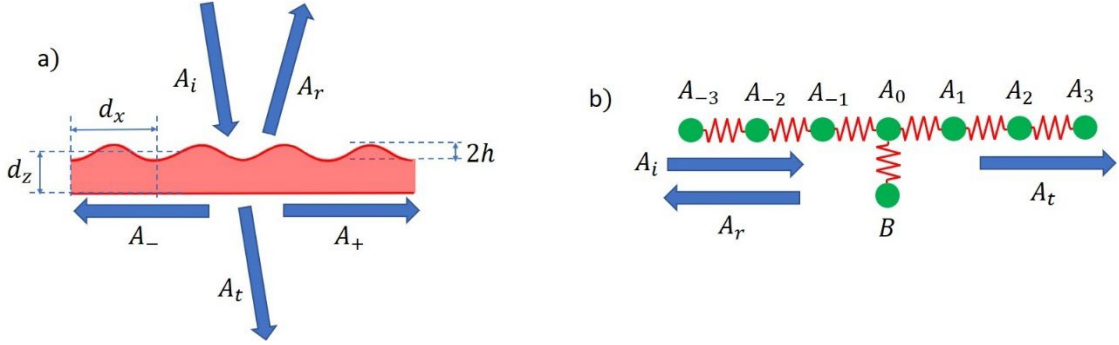

***Fig.1***. *a) Thin film with periodically corrugated surface, excited by an incident wave at a nearly normal direction; b) oscillator chain, where the oscillator $A_0$ mimics the Fabry-Perrot mode of the resonator, and the oscillator B - the planar waveguiding mode in a).*

We first derive the universal model, find its analytical solutions, then we simplify and analyze it in different limits. We present its specific solutions, corresponding to different modifications of the scheme in Fig.1.a: film with amplification, non-Hermitical films, and others. Finally, we generalize the theory to several Fano resonances. This case is not frequent in usual systems with Fano resonances, however it is very important in Fano resonances in thin films Fig.1.b: there is always a coalescence of two resonances between left- and right-propagating modes at the zero incidence angle, as well as of resonances of different order of modes. Finally, we compare a particular solution of our simple (but universal) model with the numerical solution obtained by standard numerical procedures (commercial program), in order to justify the correctness of our simple model.

**II. Mathematical description**

The equation for the dynamics of oscillators in a chain at $n \neq 0$ is:

$$\frac{\partial^2 A_n}{\partial t^2} = \omega_A^2 (A_{n-1} + A_{n+1} - 2A_n) \tag{1.a}$$

This chain supports the right and left propagating excitation waves $A_n = A_+(k)e^{ikn-i\omega t} + A_-(k)e^{-ikn-i\omega t}$, with the dispersion relation directly following from (1.a): $\omega(k)^2 = \omega_A^2(2 - e^{ik} - e^{-ik}) = 2\omega_A^2(1 - cos(k))$. $k$ is here adimensional. This curve of dispersion in a continuum limit (for $k \ll 1$) becomes straight line $\omega = \pm\omega_A k$. $\omega_A$ has also a meaning of the phase velocity of the wave in terms of lattice sites per time unit.

The central oscillator of the chain $A_0$, additionally interacts with the coupled oscillator B:

$$\frac{\partial^2 A_0}{\partial t^2} = \omega_A^2(A_{-1} + A_{+1} - 2A_0) - 2\gamma_A \frac{\partial A_0}{\partial t} + c_{AB}\omega_B(B - A_0) \tag{1.b}$$

$$\frac{\partial^2 B}{\partial t^2} = -\omega_B^2 B - 2\gamma_B \frac{\partial B}{\partial t} + c_{BA}\omega_B(A_0 - B) \tag{1.c}$$

We consider a general case. For instance, we introduced the losses $\gamma_A$ and $\gamma_B$ for the coupling and coupled oscillators $A_0$ and B. This corresponds to the losses, or gain of the Fabry-Perrot (FP) resonator and of waveguiding mode of the film, as, for instance, encountered in a scheme studied in [23]. The coupling coefficients $c_{AB}$ and $c_{BA}$ are in general not equal (more precisely not complex conjugated $c_{AB} \neq c_{BA}^*$) as it were in Hermitian case. This covers recently addressed situation of non-Hermitian thin films [22]. Note that the coupling forces are defined as $c_{AB}\omega_B(B-A_0)$ and $c_{BA}\omega_B(A_0-B)$ to make the dimensionality of the coupling constant the same as of the losses [$time^{-1}$]. And finally, the decay coefficients $\gamma_{A,B}$ can be negative, indicating the gain in the thin film.

### III. Solutions

We consider separately: i) the incident/reflected waves on the left part of chain $n < 0$: $A_n = e^{ikn-i\omega t} + re^{-ikn-i\omega t}$, where the amplitude of incident wave is normalized to unity, and $r$ is the unknown, complex valued, reflection coefficient; ii) the transmitted wave on the right $n > 0$: $A_n = te^{ikn-i\omega t}$ with the unknown transmission coefficient $t$; iii) the coupling oscillator of the chain at $n = 0$: $A_0 = A_c e^{-i\omega t}$ with the unknown amplitude $A_c$; iv) the coupled oscillator: $B = B_0 e^{-i\omega t}$ with unknown amplitude $B_0$ as well.

We write down the (1.a) in stationary regime $\frac{\partial^2 A_n}{\partial t^2} \to -\omega^2 A_n$ at points $n = -1$, and $n = 1$ also (1.b) and (1.c) at point $n = 0$. This generates four equations relating four unknowns, t, r, $A_c$ and $B_0$.

$$-\omega^2(e^{-ik} + re^{ik}) = \omega_A^2(A_c + e^{-2ik} + re^{2ik} - 2e^{-ik} - 2re^{ik}) \tag{2.a}$$

$$-\omega^2 te^{ik} = \omega_A^2(A_c + te^{2ik} - 2te^{ik}) \tag{2.b}$$

$$-\omega^2 A_c = \omega_A^2(e^{-ik} + re^{ik} + te^{ik} - 2A_c) + 2i\gamma_A \omega A_c + c_{AB}\omega_B(A_c - B_0) \tag{2.c}$$

$$-\omega^2 B_0 = -\omega_B^2 B_0 + 2i\gamma_B \omega B_0 + c_{BA}\omega_B(B_0 - A_c) \tag{2.d}$$

(2.a) and (2.b) directly leads to $A_c \equiv r + 1$, and $A_c \equiv t$. This is first important conclusion, universally relating transmission and reflection: $t - r = 1$. The usual conservation relation $|r|^2 + |t|^2 = 1$, does not hold in general, due to losses/gain in the system. In the absence of gain/losses, $\gamma_A = \gamma_B = 0$ the usual conservation relation $|r|^2 + |t|^2 = 1$ holds as well, as we derive below.

(2.d) allows to calculate the response (or gain-) function $g(\omega)$:

$$B_0 = \frac{c_{BA}\omega_B A_c}{\omega^2 - \omega_B^2 + 2i\gamma_B \omega + c_{BA}\omega_B} = g(\omega)A_c \tag{3.a}$$

From the remaining (2.c) we obtain:

$$A_c = \frac{2\omega v}{ic_{AB}\omega_B(g(\omega)-1) + 2\gamma_A \omega + 2\omega v} \tag{3.b}$$

Here $v = d\omega/dk$ is the group velocity of the wave.

Note, that (3) does not contain explicitly the wavenumber k, nor $\omega_A$. It is solely defined by the frequency and the slope of the frequency dispersion, which is the group velocity. This means that (3), and subsequent relations are applicable for the systems with arbitrary dispersion relation.

The (3) is the central solution of our study.

### IV. Simplifications

Close to the resonance, $|\omega - \omega_B| \ll \omega$, also in weak gain/loss limit $|\gamma_{A,B}| \ll \omega$, the solution (4.c) simplifies to:

$$g(\omega) = \frac{c_{BA}}{2(\omega-\omega_B) + 2i\gamma_B + c_{BA}} \tag{4.a}$$

$$A_c = \frac{2v}{ic_{AB}(g(\omega)-1) + 2\gamma_A + 2v} \tag{4.b}$$

Eliminating $g(\omega)$ from (4.a) we obtain:

$$A_c = \frac{v(2(\omega-\omega_B) + 2i\gamma_B + c_{BA})}{-ic_{AB}((\omega-\omega_B) + i\gamma_B) + (\gamma_A + v)(2(\omega-\omega_B) + 2i\gamma_B + c_{BA})} \tag{5}$$

Note that the theory (until Eq.(5)) is applicable in non-Hermitian (including PT-symmetric) cases, $c_{AB} \neq c_{BA}^*$, [22, 29-32] as well. In the following simplifications we will restrict to Hermitian coupling case.

### IV.1. Hermitian and Conservative case

In Hermitian and conservative case $c_{BA} = c_{AB}$, $\gamma_A = \gamma_B = 0$ the reflection/transmission further simplifies to:

$$t = \frac{2(\omega-\omega_B)+c_{AB}}{c_{AB}+(2-ic_{AB}/v)(\omega-\omega_B)} \qquad r = t - 1 = \frac{ic_{AB}(\omega-\omega_B)/v}{c_{AB}+(2-ic_{AB}/v)(\omega-\omega_B)} \qquad (6)$$

This predicts, that the reflection $r$ is zero at $\omega = \omega_B$, and the transmission $t$ is zero is at $\omega = \omega_B - c_{AB}/2$. This shift between reflection and transmission zeroes can be considered as the full with of the resonance $\Delta\omega_0 = c_{AB}/2$.

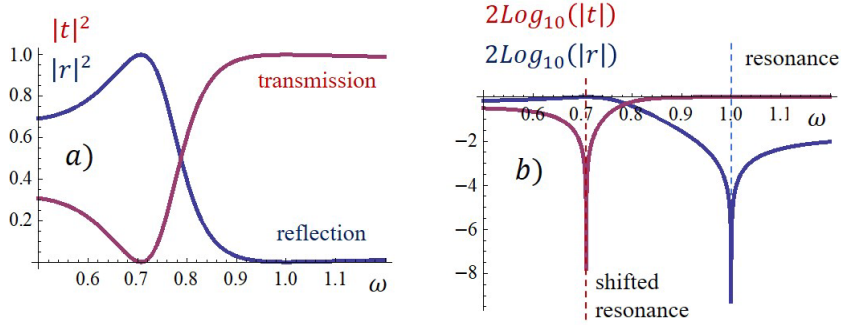

**Fig.2.** *Reflection/transmission crossing the resonance in linear a) and logarithmic (b) presentation, as obtained from (6). The reflection/transmission curve shows asymmetric form, typical for Fano scattering, which is better visible in logarithmic representation. The reflection is zero at the resonance (in this case at $\omega = \omega_B = 1$), and the peak of reflection is shifted towards smaller frequencies $\omega$ by approximately $c_{AB}/2$. The transmission maximum and zero corresponds to reflection zero and maximum respectively. The parameters: $\gamma_A = \gamma_B = 0$. $c_{AB} = c_{AB} = 0.5$.*

### IV.2. Presence of gain

In the presence of gain, one should consider (5), but one can also use (6) with the complex-valued resonance frequency $\omega_B \to \omega_B + i\gamma_B$. The calculations of a resonance lead to:

$$\omega_{im} = -\gamma_B - \frac{c_{AB}^2/v}{4+c_{AB}^2/v^2} \qquad (7)$$

Interestingly for $\gamma_B < 0$ the imaginary part of the frequency can become positive. More precisely, it becomes positive for $\gamma_B < -vc_{AB}^2/(4v^2 + c_{AB}^2)$. This indicates that the oscillator with gain starts generating. The full solution of a system consists of the solution of a driven system (5), plus the solution of the autonomous part of equation system, which is exponentially growing or decaying. A modification of above theory allows to find the solution of the autonomous equation (setting the injection equal to zero in (1)). We, however, could not get a compact analytical expression for the autonomous part of the solution, therefore we do not present it here.

The Fig.3. shows the driven solutions just below the generation threshold. Both, reflection and transmission strongly increase at the vicinity of their resonance. Also the reflection/transmission zeroes of, found in conservative cases, do not exist any more.

A simple explicit estimation of the reflection coefficient at its (shifted) resonance $\omega_R = \omega_0 - \frac{2c_{AB}}{4+c_{AB}^2/v^2}$ can be derived in the limit $c_{AB} \ll 1$ and $|\gamma_B| \ll 1$:

$$r = -\frac{1}{1+4v\gamma_B/c_{AB}^2} \tag{8}$$

This is a universal relation. In conservative case $r = -1$, as can be expected. Approaching the generation threshold $\gamma_T = -c_{AB}^2/(4v)$ the reflection at resonance tends to infinity. Expansion of the gain coefficient $\gamma_B$ and frequency $\omega$ around the singular point $(\gamma_T, \omega_R)$, $\gamma = \gamma_T + \Delta\gamma$, $\omega = \omega_R + \Delta\omega$, results in:

$$r = -\frac{c_{AB}^2/v^2}{(2i+c_{AB}/v)^2} \frac{1}{\Delta\gamma - i\Delta\omega} \tag{9}$$

which allows to estimate the width of the resonance: $\Delta\omega \approx |\Delta\gamma|$. Approaching the singular point the width of the resonance (both in reflection and transmission) approaches zero

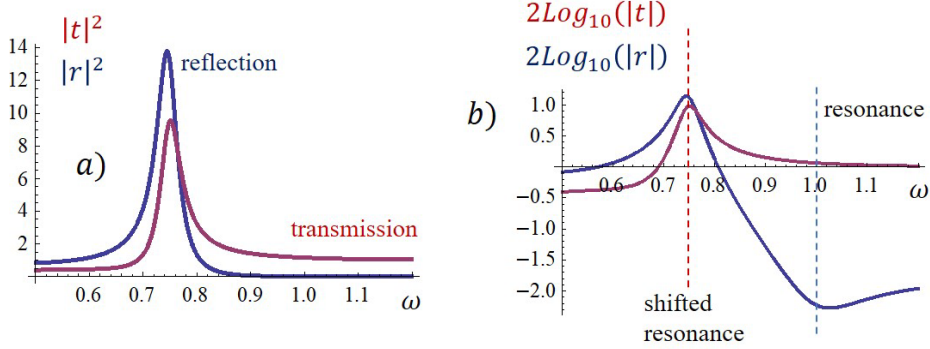

***Fig.3***. *Reflection/transmission crossing the resonance in the presence of gain $\gamma_B < 0$. The gain value is below the generation threshold. The parameters: $\gamma_A = 0$. $\gamma_B = -0.075$, $c_{AB} = c_{BA} = 0.5$, $\omega_0 = 1$*

### IV.3. Multiple Fano resonances

The theory can be generalized to the case of multiple Fano resonances. The central oscillator in the chain is then to be coupled to several oscillators $B_j$, $j = 1,2,3, ...$:

$$\frac{\partial^2 A_0}{\partial t^2} = \omega_A^2(A_{-1} + A_{+1} - 2A_0) - 2\gamma_A \frac{\partial A_0}{\partial t} + \sum_{j=1,2,3,...} c_{AB_j}\omega_j(B_j - A_0) \tag{10.a}$$

$$\frac{\partial^2 B_j}{\partial t^2} = -\omega_j^2 B_j - 2\gamma_j \frac{\partial B_j}{\partial t} + c_{BA_j}\omega_j(A_0 - B_j) \tag{10.b}$$

The above solutions for single Fano resonance (3) can be straightforwardly generalized. The amplitudes of multiple Fano resonances are calculated from (2.d):

$$B_j = \frac{c_{BA_j}\omega_j A_c}{\omega^2 - \omega_j^2 + 2i\gamma_{B_j}\omega + \omega_j c_{B_jA}} = g_j(\omega) A_c \tag{11.a}$$

And the reflection/transmission coefficients are calculated from the (2.c) modified for multiple resonances:

$$t = A_c = \frac{2\omega v}{i\sum_j c_{AB_j}(g_j(\omega)-1) + 2\omega v} \tag{11.b}$$

(11) contains the multiple resonances. The reflection follows from the relation $t - r = 1$.

An example of two resonances is shown in Fig.4. The characteristic shapes of the resonances remain as calculated for single resonance case, however quantitatively the resonance shifts are influenced by the interaction. The resonances are moderately shifted one from another (by the value comparable by the coupling constant $c_{AB} = c_{AC}$).

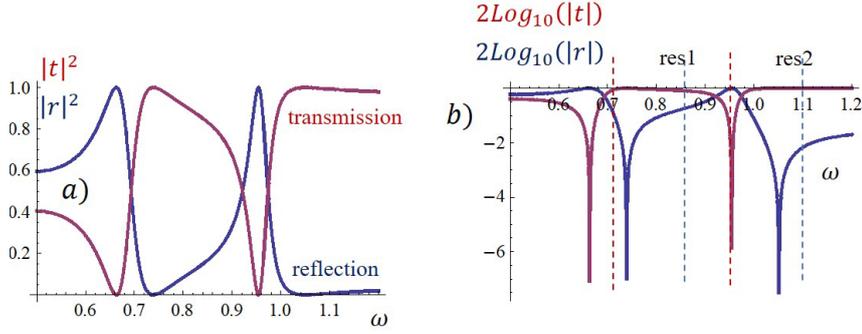

**Fig.4**. Reflection/transmission in the presence of two Fano resonances. Hermitian case. Other parameters: $\gamma_A = \gamma_B = \gamma_C = 0$. $c_{AB} = c_{AC} = 0.3$, $\omega_B = 0.86$, $\omega_C = 1.1$.

## V. Relation with the real systems

Finally, we discuss the relation between the parameters and variables in our model of oscillator chain, with those in Fig.1.a, the modulated thin films.

The planar structure (unperturbed by modulation) supports the waveguiding modes with the propagation wavenumber $k_m$ for a given frequency $\omega_0 = ck_0$ (here $k_0 = 2\pi/\lambda$ is the wavenumber of incident light in vacuum). The resonant coupling to the right/left propagating modes occurs for the incidence angles $\theta$ such, that $k_0 \sin(\theta) \pm q_x = \pm k_m$. (Here modulation wavenumber is $q_x = 2\pi/d_x$ and the period of the modulation of the surface of the film $d_x$). The propagation wavenumbers of the planar modes $k_m = \sqrt{k_0^2 n^2 - q^2}$, depend on the transverse wavenumber of the planar mode $q$, which does not have explicit analytic expression, rather obeys the transcendental relations:

$$tan(qd_z/2) = \sqrt{\frac{k_0^2}{q^2}(n^2 - 1) - 1} \qquad (12)$$

which is valid for the odd modes. For the even modes $tan(qd_z/2)$ is to be substituted by $-cot(qd_z/2)$. $d_z$ is the film thickness. The (12) holds for TE mode (or s-mode), with the vector of electric field directed perpendicularly to the plane of planar waveguide; for TM mode (14) is slightly modified, however we will not consider this case here.

In the limit of infinitely deep potential well, when the refraction index of the material of the film is infinitely large in (12), the transverse wavenumber $q$ has a simple expression, independent of polarization.

$$q \approx \frac{2\pi m}{d_z} \qquad (13)$$

Which is now valid both for the odd: $m=1,3,5,...$ and even: $m=2,4,6,...$ modes. In virtue of: $(k_0 \sin(\theta) \pm q_x)^2 = k_0^2 n^2 - q^2$, the resonant "cross" pattern is obtained in the parameters space of the incidence angle and wavenumber $(\theta, \lambda)$ for each order of the planar modes and for each polarization, where the signs $\pm$ attribute to the left/right inclined resonance lines, or, equivalently, the left/right propagating planar mode.

Another important parameter the coupling coefficient between the incident radiation and the waveguiding mode. We can estimate the coupling coefficient analytically for the harmonically modulated interface between two materials with refraction indices $n_1$ and $n_2$: $Z(x) = h \cos(q_x x) = h(e^{-iq_x x} + e^{iq_x x})/2$. Then, for instance, the amplitude of the transmitted wave reads:

$$A_t(x) = A_0 e^{ikZ(x)(n_1-n_2)} \frac{2n_1}{n_1+n_2} \approx A_0 \frac{2n_1}{n_1+n_2} + A_0 \frac{4\pi h}{\lambda} \frac{n_1-n_2}{n_1+n_2}(e^{-iq_x x} + e^{iq_x x}) + \cdots \qquad (14)$$

The series expansion in (14) assumes shallow modulation approximation: $h \ll \lambda$. This straightforwardly leads to the first order diffraction coefficient:

$$d \approx \frac{4\pi h}{\lambda} \frac{n_1-n_2}{n_1+n_2} \qquad (15)$$

and eventually to the coupling between the incoming radiation with the waveguiding mode:

$$c_{AB} \approx \frac{2h}{d_m} \frac{n_1-n_2}{n_1+n_2} \qquad (16)$$

$d_m$ is the characteristic width of the excited mode, which depends on the shape of the individual mode. Typically, it is of the order of $d_z$, but for $d_z \ll \lambda/2$ it can be larger than $d_z$ and $\lambda/2$. $c_{AB}$ is adimensional coupling, normalized to the carrier frequency, which corresponds to that one used in analytics above.

We present the rigorously numerically calculated map in Fig.5, together with the relevant cross-sections at a particular nonzero angle. (At zero angle two resonances coincide.) The estimation of $c_{AB}$ from the parameters used in the calculations (see caption of Fig.5) is $c_{AB} \approx 0.2$, which gives the width of the resonance $\Delta\omega_0 = c_{AB}/2$, as well as the displacement between the reflection and transmission zeroes equal to $\approx 0.1$, which corresponds well to these in cross-sections of Fig.5. The transmission has been calculated using our simplified model (10). The widths of the resonances correspond well between numerics and analytics. A difference in the rigorously calculated model is the background transmission, which possibly is affected by the low quality Fabri-Perrot resonator, whereas the background intensity in our analytical model is unity.

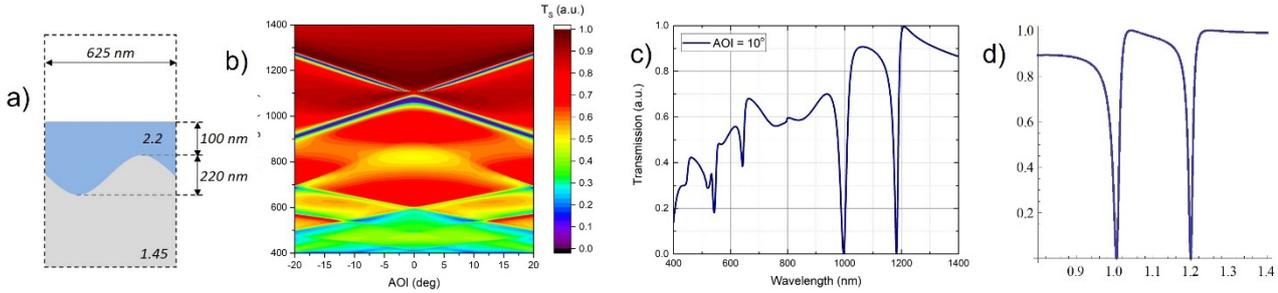

***Fig.5.*** *a) Geometry of the structure. b) transmission map in space of incidence angle and wavelength $(\theta, \lambda)$ as numerically calculated by Lumerics. c) vertical cross-section at the incidence angle 10 deg. d) corresponding transmission dependence as obtained by simplified model (15) with the coupling coefficients $c_{AB} = c_{AC} = 0.2$. The parameters are as indicated in a): $d_x = 0.625 \ \mu m$, $d_z = 0.210 \ \mu m$, $2h = 0.220 \ \mu m$, and $n = 2.2$ of the film, and $n = 1.45$ of the substrate.*

## Conclusions

Concluding, we derived analytical expressiosn for transmission/reflection coefficients of a wave propagating along the chain from a coupled oscillator. The results can be applied for the systems with Fano resonances, such as micro-ring resonator coupled to a waveguide, or a thin film coupled to near-to-normal incident radiation. This article was written basically to interpret the latter case, the reflection/transmission from/through a thin film at a near-to-normal incidence.

The model uses several phenomenological parameters, such as:

- Frequency of resonance, $\omega_B$;
- Coupling constants $c_{AB}, c_{BA}$;
- Decay rates $\gamma_A, \gamma_B$, which can represent gain for $\gamma_A, \gamma_B < 0$
- Group velocity of the propagating modes $v$;

The derived formulas allow to estimate:

- The transmission/reflection function (3), and its simplifications (4-6)
- The frequency of resonance in reflections (maximum reflection) $\omega_{rez} = \omega_B - c_{AB}/2$; The maximum transmission (zero reflection) frequency remains at resonance $\omega_B$.
- The separation between transmission ar reflection zeroes, which is an estimation of the width of the resonance in conservative case $\Delta\omega_0 = c_{AB}/2$;
- In case of oscillators with gain the singular point occurs at $\gamma_T \to -c_{AB}^2/4$, when the field starts to grow exponentially. Close to the instability point $\gamma_T = -c_{AB}^2/(4v)$: the width of the resonance with amplification is equal to the distance frim generation point $\Delta\omega \approx |\Delta\gamma|$,

In addition, the above characteristics for multiple Fano resonances have been derived as well (15).

**Acknowledgement:**

The study is supported by Spanish Ministry of Science, Innovation and Universities (MICINN) under the project PID2022-138321NB-C21, also by the Research Council of Lithuania (LMTLT), under agreement No. S-LT-TW-24-9. The author thanks to L. Grineviciute, J. Nikitina, and L.Letelier for numerical simulations with Lumerics.